\documentclass[a4paper]{panl}
\usepackage{cite}
\usepackage{wrapfig}
\usepackage{graphicx}
\usepackage{amssymb}
\usepackage{amsfonts}
\usepackage{amsmath}
\usepackage{longtable}
\usepackage{rotating}
\usepackage{lscape}
\usepackage{epsfig}
\usepackage{multirow}
\usepackage[usenames]{color}

\usepackage{bm}

\def\1#1{{\bf #1}}

\originalTeX

\begin{document}

\issuearea{Physics of Elementary Particles and Atomic Nuclei. Theory}

\title{Threshold Collision Energy of the QCD Phase Diagram  Tricritical Endpoint 
}

\maketitle
\authors{
K.\,A.\,Bugaev$^{a}$\footnote{E-mail: bugaev@fias.uni-frankfurt.de}, 
R.\,Emaus$^{b}$,
V.\,V.\,Sagun$^{a,c}$\,
A.\,I.\,Ivanytskyi$^{a}$,
L.\,V.\,Bravina$^{b}$,
}
\authors{
D.\,B.\,Blaschke$^{d,e,f}$,
E.\,G.\,Nikonov$^{g}$,
A.\,V.\,Taranenko$^f$,
E.\,E.\,Zabrodin$^{b,f,h}$,
}
\authors{
G.\,M.\,Zinovjev$^{a}$
}
\from{$^{a}$\,Bogolyubov Institute for Theoretical Physics, Metrologichna str. 14$^B$, Kiev 03680, Ukraine}
\vspace{-3mm}
\from{$^{b}$\,Department of Physics, University of Oslo, PB 1048 Blindern, N-0316 Oslo, Norway}
\vspace{-3mm}
\from{$^{c}$\,CENTRA, Instituto Superior T$\acute{e}$cnico, Universidade de Lisboa, Av. Rovisco Pais 1, 1049-001 Lisboa, Portugal}
\vspace{-3mm}
\from{$^{d}$\,Institute of Theoretical Physics, University of Wroclaw, pl. M. Borna 9, 50-204 Wroclaw, Poland}
\from{$^{e}$\,Bogoliubov Laboratory of Theoretical Physics, JINR Dubna, Joliot-Curie str. 6, 141980 Dubna, Russia}
\from{$^{f}$\,National Research Nuclear University ``MEPhI'' (Moscow Engineering Physics Institute), Kashirskoe Shosse 31, 115409 Moscow, Russia}
\from{$^{g}$\,Laboratory for Information Technologies, JINR, Joliot-Curie str. 6, 141980 Dubna, Russia}
\from{$^{h}$\,Skobeltzyn Institute of Nuclear Physics, Moscow State University,
119899 Moscow, Russia}

\begin{abstract}
Using the most advanced  formulation of the hadron resonance gas  model we analyze the two sets of 
irregularities  found at chemical freeze-out of central nuclear-nuclear collisions at   the center of mass  energies 3.8-4.9 GeV and 7.6-9.2 GeV.  In addition to previously reported irregularities at the collision energies 4.9 GeV and 9.2 GeV  we found  sharp peaks of baryonic charge density.  Also we analyze the collision energy dependence of the modified Wroblewski factor and the strangeness suppression factor. 
Based on the thermostatic properties of the mixed phase  of a 1-st order phase transition and  the ones of the Hagedorn mass spectrum   we explain, respectively, the reason of observed chemical equilibration of strangeness   at the collision energy  4.9 GeV and above 8.7 GeV.   It is argued that the both sets  of irregularities  possibly evidence for two phase transitions, namely, the 1-st order transition at 
lower energy range and the 2-nd order transition at higher one.  In combination with a  recent analysis of  the light nuclei  number fluctuations we conclude that the center of mass  collision energy range  8.8-9.2 GeV may be in  the 
nearest vicinity of  the  QCD tricritical endpoint.  The properties of the 
phase existing between two phase transitions are revealed and discussed. 
\end{abstract}
\vspace*{6pt}

\noindent
PACS: 25.75.$-$q, 25.75.Nq


\bigskip 

\label{sec:intro}
\section*{Introduction}

Theoretical and experimental searches for the (tri)critical endpoint ((3)CEP)  of  the
quantum chromodynamics  (QCD) phase diagram form one of  the most important directions of  physics of heavy ion collisions \cite{Misha2,WP9}. 
Despite the claims that the QCD CEP  parameters  can be  extracted from  the RHIC data \cite{Roy1}, the real situation  is not so clear \cite{antiRoy1}. 
In part, such a situation exists because the rigorous and reliable definition of finite volume analog of   (3)CEP does  not exist.  
In our opinion, therefore, the question is not whether the work \cite{FSS} can justify the finite size scaling 
for the isothermal  compressibility  as it is assumed in  \cite{Roy1} or for the isentropic compressibility as its is argued in  \cite{antiRoy1},  but whether there is sufficient experimentally established signals  of the (3)CEP  formation.
Furthermore, we are sure that  to safely locate the threshold collision energy of the (3)CEP one
needs a  simultaneous match of  thermodynamical, statistical and hydrodynamical  signals of its formation. 

Recently the thermodynamical and hydrodynamical signals of the mixed quark-gluon-hadron  phase formation 
were revealed  \cite{Bugaev:2014,Bugaev:2015}  from the analysis of different irregularities observed at chemical freeze-out (CFO)  \cite{SFO,Veta14,Bugaev:2016a}. Moreover, such signals  were observed  at two ranges of the center-of-mass collision energies, namely $\sqrt{s_{NN}} \simeq 3.8-4.9$ GeV and  $\sqrt{s_{NN}} \simeq  7.6-9.2$ GeV  \cite{Bugaev:2014,Bugaev:2015, Bugaev:2016a}.  The thermodynamic signals include two sharp peaks of the  trace anomaly $\delta = \frac{(\epsilon - 3p)}{T^4}$ (here $\epsilon$, $p$ and  $T$ denote, respectively, the energy density of the system, its pressure and temperature) observed at  $\sqrt{s_{NN}} = 4.9$ GeV and 
$\sqrt{s_{NN}} = 9.2$ GeV, while the hydrodynamic  signals consist of 
the highly correlated quasi-plateaus in the collision energy dependence 
of the entropy per baryon, total pion number per  baryon, and thermal pion number per baryon 
which were found  at the two  ranges of the center-of-mass collision energies discussed above  
\cite{Bugaev:2014,Bugaev:2015}.  Thermodynamical and hydrodynamical signals
in the low collision energy region can be easily interpreted  since they can 
be  related with each other via the generalized shock adiabat model  
\cite{GSA1,GSA2}
which shows that  there is one-to-one correspondence between the peak of  $\delta$  at CFO and the peak of  $\delta$  at the shock adiabat \cite{Bugaev:2014,Bugaev:2015}. Moreover,  at CFO the low energy signals are accompanied  by the strong jumps of the pressure $p$ and  the  effective number of degrees of freedom $p/T^4$ \cite{Bugaev:2014,Bugaev:2015,SFO}. In contrast to these findings, the high energy signals are less pronounced and, hence, their 
interpretation is not straightforward  \cite{Bugaev:2014,Bugaev:2015,Bugaev:2016a} and, therefore,  one needs additional analysis of  the collision energy range $\sqrt{s_{NN}} \simeq  7.6-9.2$ GeV. 

In a recent work on the analysis of critical fluctuations of light nuclei \cite{Ko2017} at CFO  it was argued  that at the collision energy $\sqrt{s_{NN}} = 8.8$ GeV the QCD matter has strongest fluctuations and, hence, during its evolution  this matter passes
through the CEP.   This work  puts forward a strong argument in favor of the hypothesis that  at the vicinity of the point  
$\sqrt{s_{NN}} = 8.8$ GeV, i.e. almost at the location of  the second peak of  trace anomaly $\delta$, there occurs another phase transformation. Moreover, the work \cite{Ko2017}  concludes that  this is a vicinity of CEP, i.e. it has to be a 2-nd order phase transition.  Thus, in addition to the thermodynamic and hydrodynamic signals of  a possible  phase transformation at  the collision energy range $\sqrt{s_{NN}} \simeq  7.6-9.2$ GeV  \cite{Bugaev:2014,Bugaev:2015,Bugaev:2016a}    the work \cite{Ko2017},
first of all,  gives an additional and independent evidence  in  favor of such a transformation and, moreover,  it  
provides us with the necessary  statistical  (fluctuation) signal. 
However, in combination with the previous findings on the signals of 1-st order phase transition at the collision energies 
$\sqrt{s_{NN}} \simeq 4.3-4.9$ GeV  \cite{Bugaev:2014,Bugaev:2015, Bugaev:2016a} one should conclude that these are the signals of two phase transitions and, therefore, at the vicinity of collision energies  $\sqrt{s_{NN}} \simeq 8.8-9.2$ GeV there may exist not a CEP, but a 3CEP.  It should be stressed that this is highly nontrivial conclusion, since neither the lattice QCD nor the QCD inspired field-theoretical models can presently tell us for sure whether QCD has the CEP or 3CEP. 

Furthermore, if we observe the signals of two phase transitions there are two basic questions related to this fact: 
(I) can one determine which of these transitions is deconfinement and which one is the chiral symmetry restoration? and (II) what kind of phase can be probed at the collision energies $\sqrt{s_{NN}} \simeq 4.9-8.8$ GeV? The present work is an attempt to 
formulate the proper  answers to these two major  questions.

The work is organized as follows.  In the next section we present the details of hadron resonance gas model and the fitting procedure of hadronic multiplicities. Section 3 is devoted to an analysis of new irregularities at CFO.  In Section 4 we discuss the 
properties of explicit thermostat and apply them to explain the strangeness  equilibration   observed  at the collision energy 
$\sqrt{s_{NN}} = 4.9$ GeV. Also  in this section  we estimate the number of degrees of freedom of a phase possibly 
formed at  $\sqrt{s_{NN}} \simeq 4.9-8.8$ GeV.  The equation of state of (almost) massless particles with the relativistic treatment 
of  hard-core repulsion  is analyzed in Section 5. Our conclusions are summarized n Section 6. 

\label{sec:analysis}
\section*{Analysis of hadron multiplicities}

First we perform the cross-check of the previously  obtained results on the irregularities at CFO \cite{Bugaev:2014,Bugaev:2015, Bugaev:2016a}. For this purpose we employ the newest  version of the hadron resonance gas model (NHRGM) \cite{Bugaev:2016,Sagun17,Bugaev17} which allows us to safely  go beyond the usual Van der Waals approximation for the multicomponent case, i.e. for several different hard-core radii of hadrons.  Also below we reveal 
the new irregularities and  give their explanation based on the hypothesis of existence of  two phase transitions.

The  new key element of the NHRGM  is an inclusion of the surface tension which is generated by the hard-core 
repulsion.  Therefore, this equation of state is called the induced surface tension one. 
For  the system with multicomponent  hard-core repulsion in the grand canonical ensemble   this equation of state  is given   by the  system of two coupled equations
\begin{eqnarray}
\label{EqI}
p&=&\sum_n \, p_n^{id} \, (T, \mu_n-pV_n-\Sigma S_n) \,,\\
\label{EqII}
\Sigma&=&\sum_n \, R_n \,p_n^{id} (T, \mu_n-pV_n-\alpha\Sigma S_n) \, ,
\end{eqnarray}
where the sums  are  running  over all particles (and antiparticles)  with the chemical 
potentials  $\mu_n$, the hard-core radii $R_n$, the proper volumes $V_n=\frac{4}{3}\pi R_n^3$ and 
the proper surfaces $S_n=4\pi R_n^2$. Here $p_n^{id}(T, \mu_n)$  denotes the partial pressure 
of the point-like particles of sort $n$ with  the degeneracy $g_n$ and the mass $m_n$ which in  
case of the Boltzmann statistics is 
\begin{eqnarray}
&&p^{id}_n (T, \nu) = g_n   \int\frac{d{\bf k}}{(2\pi^3)} \frac{k^2}{3\, E_n (k)}e^{\frac{\nu - E_n (k) }{T} }\, .
\label{EqIII}
\end{eqnarray}
Here $E_n(k) = \sqrt{{\vec k}^2 + m_n^2}$ is the energy of particle with the 3-momentum $\vec k$  and $\nu$ is the effective chemical potential.

In  the grand canonical ensemble the one component   Van der Waals equation of state  (EoS) \cite{VDW1, SFO,HRGM:13} can be obtained from the 
pressure of ideal gas, classical or quantum,  by the modification  of the chemical potential as $\mu_1  \rightarrow \mu_1- 4 V_1 p$ \cite{Rischke91}.
In the one component  NHRGM   the free energy associated with the excluded volume of  particle  $ - 4 V_1 p$ is split up
  into a sum of the volume part $ - V_1 p$,
which is proportional to the pressure $p$, and the surface part  $ -  S_1 \Sigma$, which is proportional 
to the induced surface tension coefficient $\Sigma$ (here $S_1$ is the proper surface of particle having the hard-core radius $R_1$). 
Such a splitting  allows one  not only to account  for the second virial coefficient, but also to reproduce  the third and  the fourth ones of  the gas of hard-spheres of the radius $R_1$  with a good accuracy \cite{Bugaev:2016,Sagun17}.  The system (\ref{EqI}) and  (\ref{EqII}) is a generalization of  such an approach  to a multicomponent case.
The  parameter  $\alpha > 1$ switches  between the excluded-volume and the proper-volume regimes.
Numerical  analysis \cite{Sagun17} shows that for $\alpha=1.245$   the system  (\ref{EqI}) and (\ref{EqII}) correctly reproduces the one component  \cite{CSeos}  and multicomponent \cite{CSmultic} versions of the well-known Carnahan-Starling EoS up to  the packing fractions $\eta = \sum\limits_{n=1} V_n \, \rho_n \le 0.22-0.24$ (here $\rho_n$ is the particle number density of hadron species $n$).   Note that  the Carnahan-Starling EoS and its multicomponent  version \cite{CSmultic}  are well-known in the theory of simple liquids since they are able to reproduce the EoS of hard spheres up to its transition to a liquid state \cite{Liquids1,Liquids2}.
 
 The particle density of hadrons of sort $n$, 
\begin{equation}\label{EqIV}
\rho_n \equiv \frac{\partial  p}{\partial \mu_n} = \frac{1}{T} \cdot \frac{p_n \, a_{22} 
- \Sigma_n \, a_{12}}{a_{11}\, a_{22} - a_{12}\, a_{21} } \,,
\end{equation}
is expressed in terms of the auxiliary coefficients 
\begin{eqnarray}
a_{11} = 1 +  \sum_n V_n  \frac{p_n}{T} \,,  \quad 
a_{12} =  \sum_n S_n \frac{p_n}{T} \, , \\
a_{21} =  \sum_n V_n \frac{\Sigma_n}{T} \, , \quad
a_{22} = 1 +  \sum_n \alpha  S_n \frac{\Sigma_n}{T} \,.
\end{eqnarray}
At CFO  the ratio of total hadronic multiplicities of sorts $i$ and $j$  is given by
\begin{equation}
\label{EqVII}
R_{ij}\equiv\frac{N^{tot}_i}{N^{tot}_j}=
\frac{\rho_i+\sum_{l\neq i}\rho_l \, Br_{l\rightarrow i}}{\rho_j+\sum_{l\neq j}\rho_l \, Br_{l\rightarrow j}}\,.
\end{equation}
This expression accounts for the hadrons which appear in strong decays 
of resonances  with the branching ratios $Br_{l\rightarrow j}$. 
The  total chemical potential of hadron of  the sort $n$  is  $\mu_n = Q_n^B \mu_B + Q_n^S \mu_S + Q_n^{I3} \mu_{I3}$ 
and it 
depends on  the baryonic chemical potential $\mu_B$, the strange chemical potential $\mu_S$, the isospin third projection 
chemical potential $\mu_{I3}$ and the corresponding charges $Q_n^B, Q_n^S, Q_n^{I3}$ of this hadron.
More details on the NHRGM can be found in  \cite{Sagun17}.

Similarly to \cite{Bugaev:2016a}, 
here we fit the high quality experimental multiplicity ratios  measured at AGS for $\sqrt s_{NN} = 2.7, 3.3, 3.8, 4.3, 4.9$ GeV 
\cite{AGS1,AGS2,AGS2b,AGS3,AGS4,AGS5,AGS6,AGS7,AGS8}, the NA49 data measured at SPS energies  $\sqrt s_{NN}=6.3, 7.6, 8.8, 12.3, 17.3$ GeV  \cite{SPS1,SPS2,SPS3,SPS4,SPS5,SPS6,SPS7,SPS8,SPS9} and the STAR data measured  at 
RHIC energies  $\sqrt s_{NN}=9.2, 62.4, 130, 200$ GeV \cite{RHIC}.
These  experimental data allow us to construct 111 independent ratios measured 
at 14 collision energies \cite{SFO,HRGM:13}. 
In order   to verify the stability of the  results on the new signals of  phase transition  between the  hadron  and quark gluon  matters found   in   \cite{Bugaev:2016a} with the help of the  HRGM based on the Van der Waals approximation
and to  improve the description of  (anti)$\Lambda$-hyperons and  (anti)protons 
we added 10 experimental data points  of the  $\Lambda/p$  ratio (on its exclusive role see a discussion in \cite{Bugaev:2015}).   Although  the $\Lambda/p$ ratio can be expressed via other ratios, i.e. it is dependent, but   at AGS and SPS energies this ratio  has rather small error bars compared to other ratios and this is important  for the fit stability  \cite{Bugaev:2016a}.
Hence, in present work we analyze 121 ratios. 
\begin{figure}[ht]
\centerline{
\includegraphics[width=77mm]{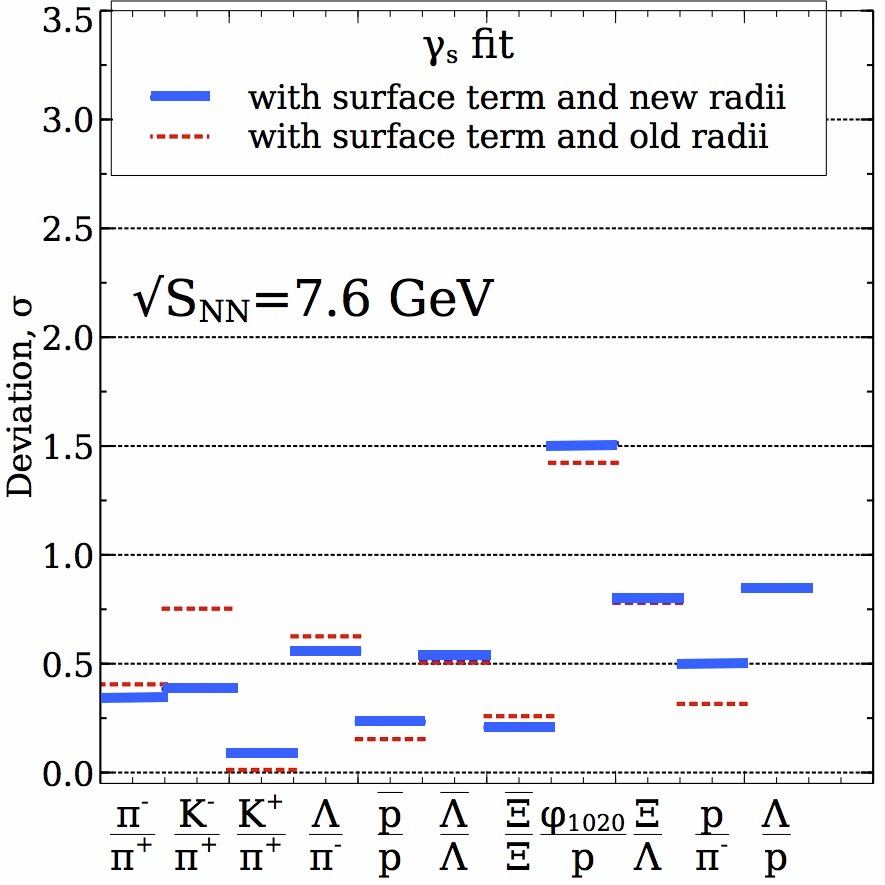}
  \hspace*{4.4mm}
\includegraphics[width=77mm]{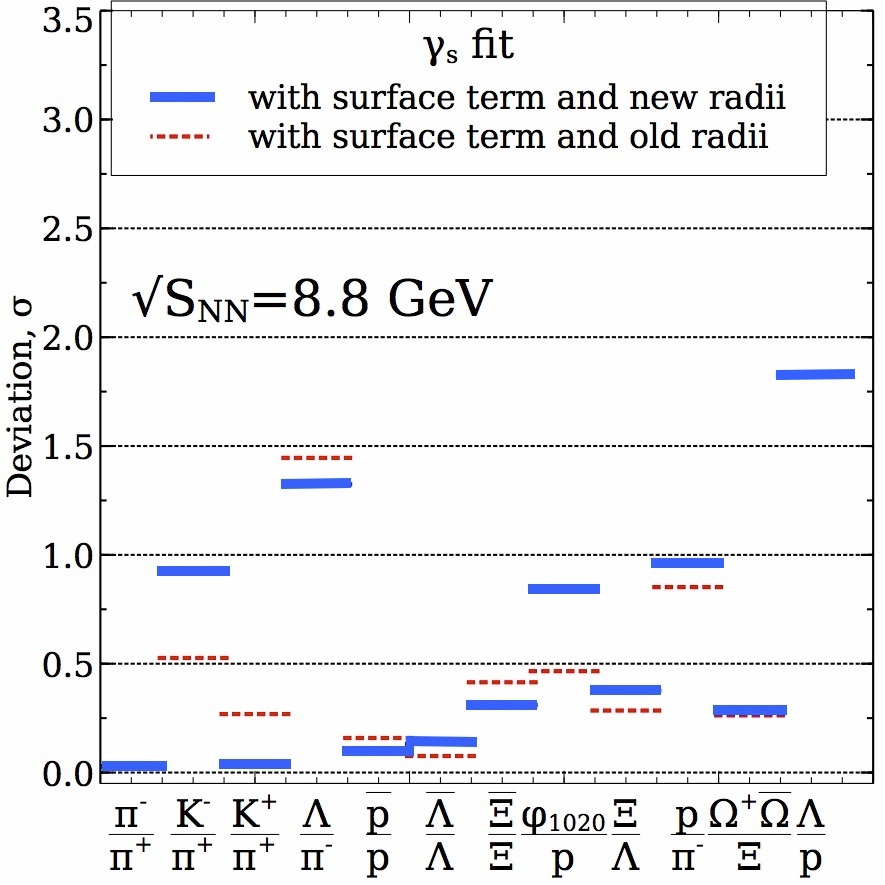}
  }
 \caption{Deviations of theoretically predicted hadronic yield ratios from experimental values in units of
 experimental error $\sigma$ are shown for the center of mass collision energies $\sqrt{s_{NN}} = 7.6$ GeV 
 and $\sqrt{s_{NN}} = 8.8$ GeV. 
 Blue lines correspond to the IST EoS fit, while the 
 red  lines correspond to the original HRGM  fit \cite{Sagun17}.}
\label{Fig1}
\end{figure}

In addition,  compared to works \cite{Bugaev:2016a,Bugaev:2016,Sagun17}  we used a new set of global fitting parameters to minimize the mean square deviation 
$\chi^2=\sum_{p=1}^{121}\frac{(R_p^{theor}-R_p^{exp})^2}{\sigma_p^2}$, where 
the experimental ratios are denoted as $R_p^{exp}$, the theoretical ones as $R_p^{theor}$ and 
 the summation is carried out over all data points with the weights defined by 
the experimental error $\sigma_p$. The  minimization of $\chi^2$  showed that compared to previous results \cite{Bugaev:2015,Bugaev:2016a,Sagun17} the values of  local fitting parameters
$T$, $\mu_B$, $\mu_{I3}$ and $\gamma_s$ are practically the same. 
We found  the best description of  the hadronic multiplicity ratios for the hard-core radius of pions
$R_\pi = 0.2$ fm,    (anti)$\Lambda$-hyperons  $R_\Lambda =0.05$ fm,
 (anti)protons $R_p = 0.37$ fm,  other baryons  $R_{b}=0.4$ fm and  other mesons  $R_{m} = 0.43$ fm.
 This set of radii, i.e. the new radii afterwards,    corresponds to  a global minimum of $\chi^2/dof = 65.42/65 \simeq 1.01$ which provides practically the same  quality of fit as the set of global fitting parameters reported in \cite{Bugaev:2016a,Sagun17}. 

\begin{figure}[htb!]
\centerline{
\includegraphics[height=7 cm]{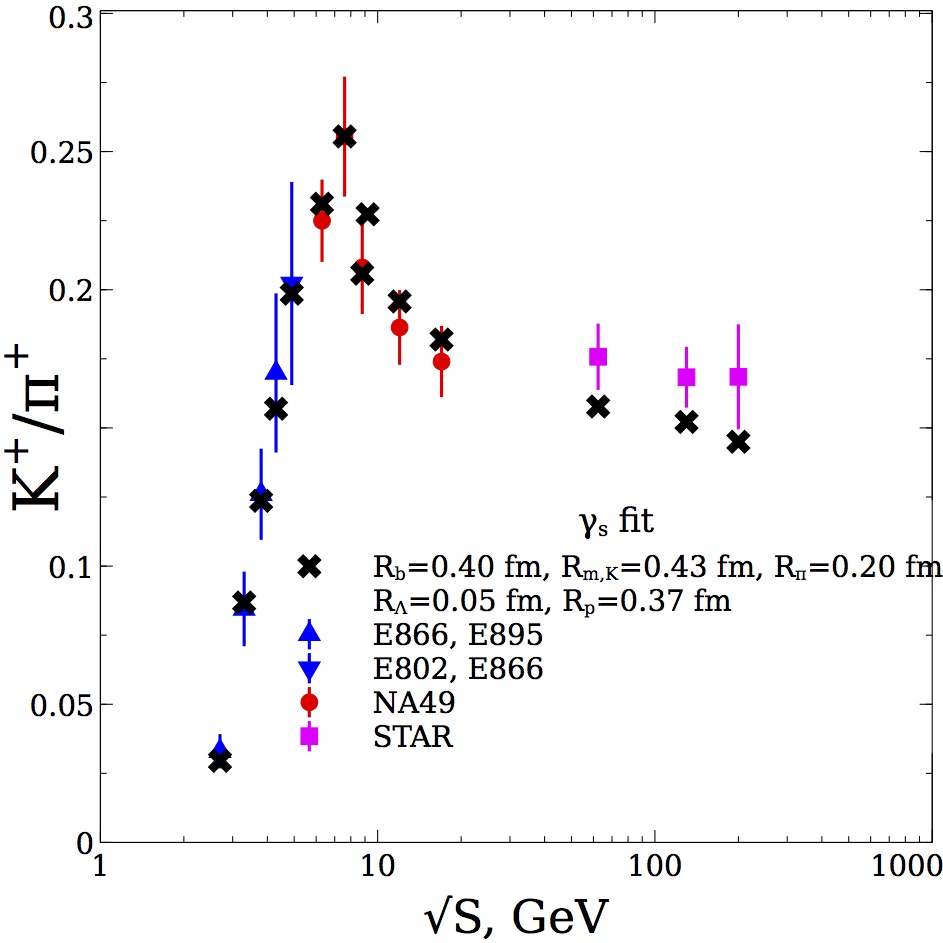} \hspace*{4.4mm}
\includegraphics[height=7.cm]{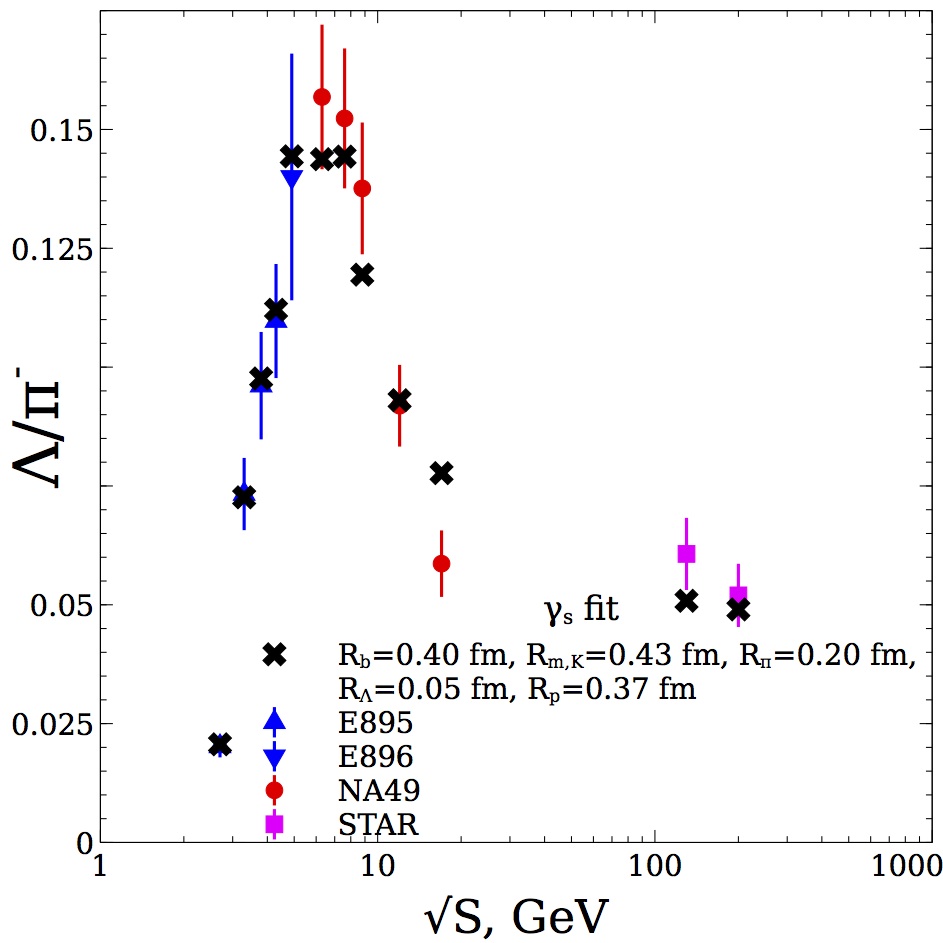}   
}
 \caption{The fit results obtained by the NHRGM with the new radii.  {\bf Left panel:}$\sqrt{s_{NN}}$  dependence  of  $K^+/\pi^+$. {\bf Right panel:}  $\sqrt{s_{NN}}$  dependence  of $\Lambda/\pi^{-}$.  
}
  \label{Fig2}
\end{figure}

The set of old radii fitted in  \cite{Sagun17} provided $\chi_1^2/dof=57.099/55 \simeq 1.038$  and it included  the
hard-core radii of  pions $R_{\pi}$=0.15 fm, kaons $R_{K}$=0.395 fm,  $\Lambda$-hyperons $R_{\Lambda}$=0.085 fm,
baryons $R_{b}$=0.365 fm and  mesons $R_{m}$=0.42 fm. Comparing these two sets of hard-core radii and their 
fit quality one concludes that  there are slight changes only which lead to little improvements. This can be seen from the 
ratios shown in Figs. \ref{Fig1} and \ref{Fig2}. Therefore, our main conclusion is that the results  previously obtained within
the multicomponent  Van der Waals EoS  are almost 
 unchanged and, hence,  there is no need to revise the signals of phase transitions
suggested earlier in  \cite{Bugaev:2014,Bugaev:2015,Bugaev:2016a}.
\begin{figure}[htb!]
\centerline{\includegraphics[width=75.2mm]{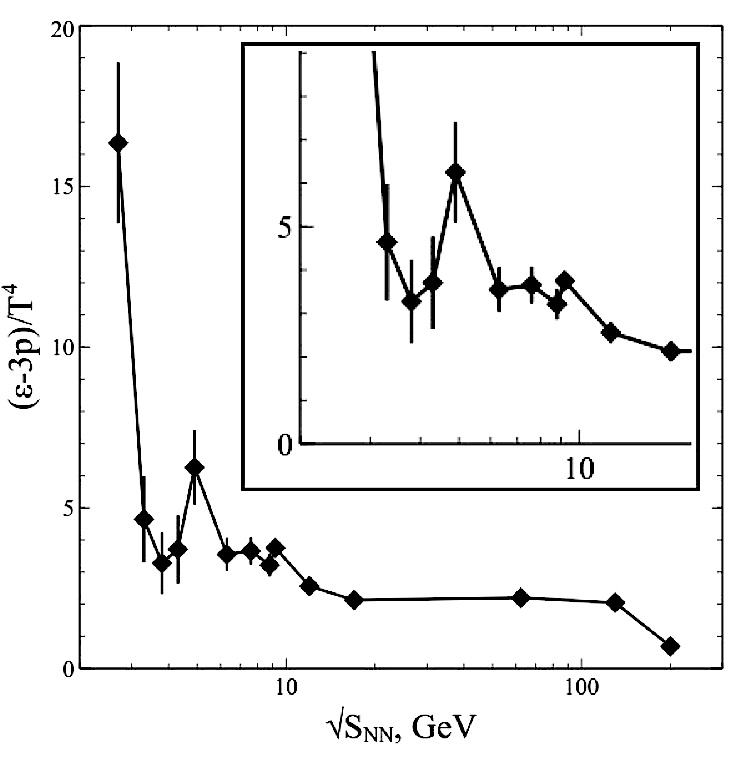}\hspace*{2.2mm}
\includegraphics[width=100mm]{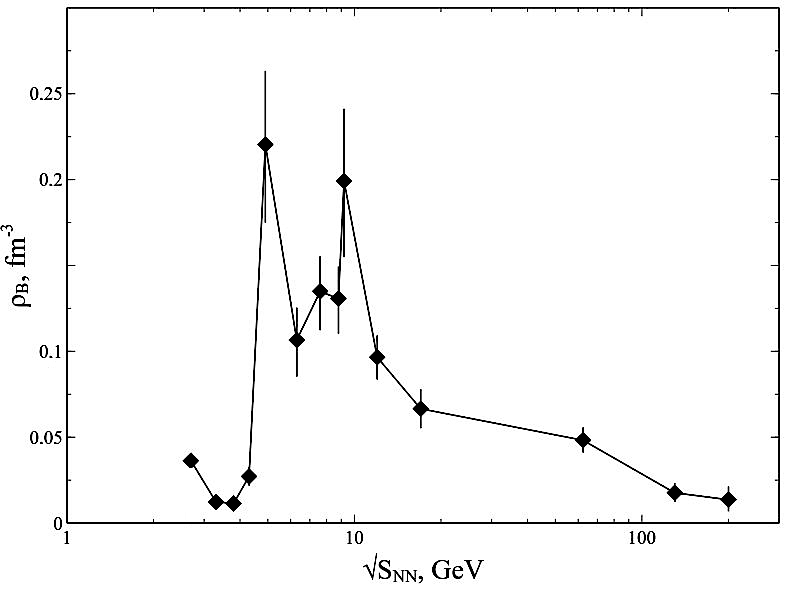}}
\caption{Collision energy dependence of the trace anomaly $\delta$ (left) and baryonic charge density $\rho_B$ (right) at CFO.
The points are connected in order to guide the eye. 
}
\label{Fig3}
\end{figure}

Concerning the small  values of the hard-core radii of pions and (anti)$\Lambda$ hyperons we would like to mention that 
these radii are an effective ones.  It is possible that small value of  the pion hard-core radius is a reflection of  necessity to account for its relativistic  nature   at  temperatures above 120 MeV \cite{RelVDW1,RelVDW2}.  
 On the other hand the small hard-core radius of   (anti)$\Lambda$ hyperons may reflect some subtleties  of their interaction with 
hadronic medium. 
Hence, we hope that their values provide an important information for microscopic models 
of hadronic interaction  which will have to explain our results. 

\label{sec:irreg}
\section*{New irregularities at CFO}

Further results concerning the trace anomaly and baryon density are shown
in Fig.~\ref{Fig3}. From the both panels of this figure one can see that 
each peak of the trace anomaly $\delta$  is accompanied by a strong peak of the total baryonic density $\rho_B = \frac{\partial p}{\partial \mu_B}$. Moreover, the  trace anomaly peak at  $\sqrt{s_{NN}} = 9.2$ GeV has a small amplitude, while  its 
counterpart in the baryonic density has a large amplitude accompanied by 
small error bars making this peak quite pronounced. 
\begin{figure}[ht]
\centerline{\hspace*{-0.44mm}
\includegraphics[width=99mm]{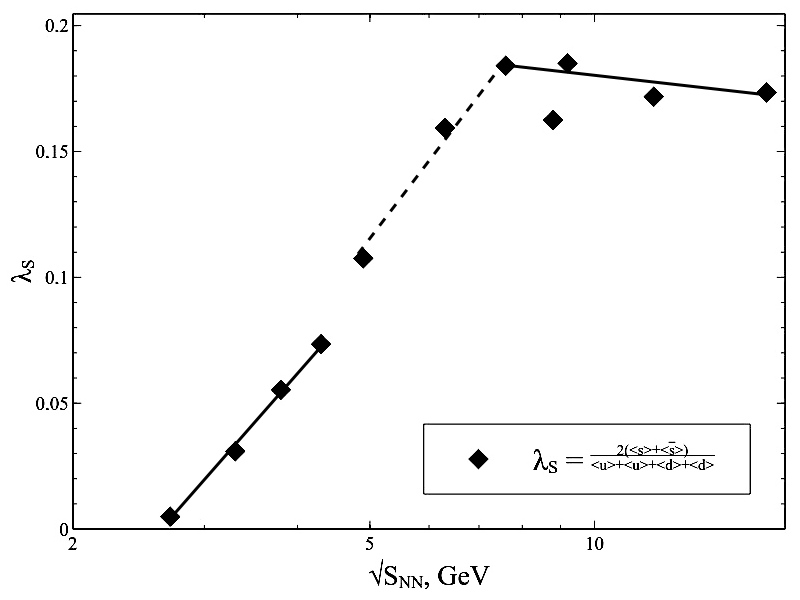}
\hspace*{2.2mm}
\includegraphics[width=77.mm]{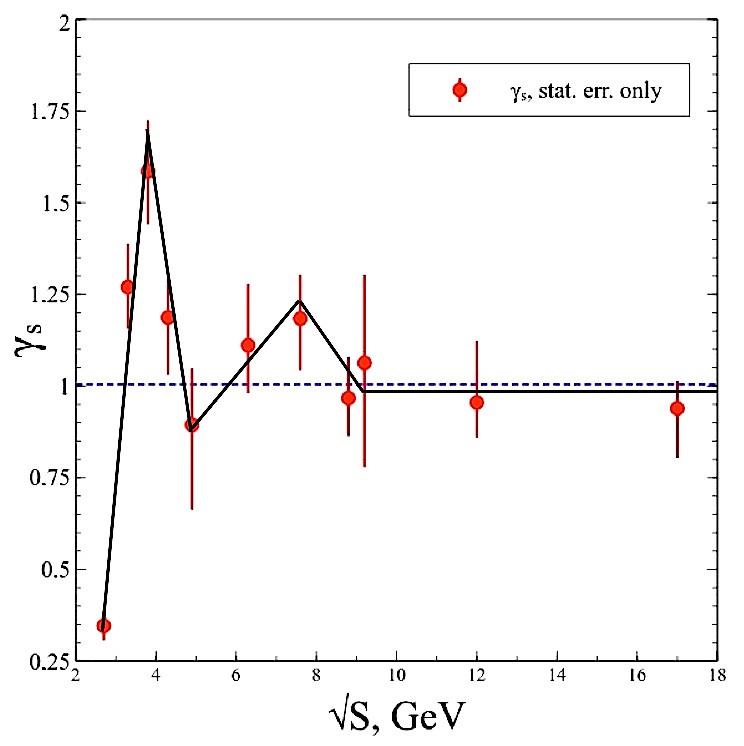}
 } 
 \caption{{\bf Left panel:} Collision energy dependence of  the modified  Wroblewski factor  $\lambda_s$.  Diamonds   show the NHRGM  description of experimental data.  At and above $\sqrt{s_{NN}}= 7.6$ GeV there is a saturation of $\lambda_s$. 
   {\bf Right panel:} Collision energy dependence of  the strangeness suppression factor $\gamma_s$ found by the NHRGM
   clearly indicates the strangeness equilibrium dale at $\sqrt{s_{NN}}= 4.9$ GeV.  
   The points are connected in order to guide the eye.
   }
  \label{Fig4}
\end{figure}

From  the left panel of  Fig.\ \ref{Fig4} one can see even more dramatic changes in the collision energy dependence  of 
the  modified Wroblewski factor $\lambda_s$ \cite{Wrobl} which we define as 
\begin{eqnarray}\label{EqVIII}
\lambda_s  \equiv \frac{2 \sum\limits_{n} (N_n^S + N_n^{\bar{S}})  \rho_n }{ \sum\limits_{n} (N_n^u +  N_n^{\bar{u}} +
N_n^d +  N_n^{\bar{d}}) \rho_n } \,,
\end{eqnarray}
where in the numerator $N_n^S$ and $N_n^{\bar{S}}$ denote, respectively, the number of strange valence quarks and antiquarks
in the hadron of sort $n$, whereas  $N_n^u$ and $N_n^d$ 
in the denominator denote, respectively, the number of $u$ and $d$ 
quarks in it 
(apparently, $N_n^{\bar{u}}$ and $N_n^{\bar{d}}$ play the same role for antiquarks). Note that the denominator in Eq. (\ref{EqVIII}) differs from the traditional Wroblewski factor \cite{Wrobl} because it accounts for a whole  number of  $u$ and 
$d$  valence quarks and antiquarks, and not only the ones which are  paired  to their valence  antiquarks (quarks).

As one can see from  Fig.\ \ref{Fig4} the factor $\lambda_s$  demonstrates 
a  jump right in the collision energy region which is associated with the mixed phase formation, i.e. for  $\sqrt{s_{NN}} =4.3-4.9$ GeV, and in addition   it  shows  a change of slope at $\sqrt{s_{NN}}  > 7.6$ GeV.  Note that  a similar behavior was found in \cite{Bugaev:2016a} for the ratio  $\frac{\Delta \Lambda}{\Delta\, p} = \frac{\Lambda - \bar \Lambda}{p - \bar p}$  and we confirm that in the present analysis  it behaves similarly.  
The found collision energy dependence of $\lambda_s$  further supports
the hypothesis of Ref.  \cite{Bugaev:2016a} that  such a  behavior of  $\frac{\Delta \Lambda}{\Delta\, p}$ and, hence,  of
  $\lambda_s$   is an indicator of two phase transformations. Indeed,  the observed jump of these ratios  is located in  the collision energy range of the mixed phase formation (i.e. it is a 1-st order phase transition), then a change of  their  slope at $\sqrt{s_{NN}} > 7.6$ GeV  can be naturally associated with a weak 1-st order or a 2-nd order phase transition. 
  
The collision energy behavior of the strangeness suppression factor $\gamma_s$ \cite{Rafelski} shown in the right panel of Fig.\ \ref{Fig4} demonstrates  more complicated dependence. 
Here one   can see two peaks for  $\sqrt{s_{NN}} < 8.8$ GeV which evidence about the  strangeness enhancement (over-saturation) and {\it the strangeness equilibrium dale} at $\sqrt{s_{NN}}= 4.9$ GeV.  At and above $\sqrt{s_{NN}} \ge  8.8$ GeV one observes a chemical equilibration of strange charge, since  
 $\gamma_s =1$. One might think that compared to the  $\gamma_s$ peak  at  $\sqrt{s_{NN}} = 3.8$ GeV the smaller one seen 
at $\sqrt{s_{NN}} = 7.6$ GeV is not statistically significant.  However, a thorough analysis of two alternative  approaches  to describe the  
chemical non-equilibrium  of strangeness performed in  \cite{Bugaev:2016ujp} 
shows that the  $\gamma_s$ peak in the vicinity 
of  $\sqrt{s_{NN}} = 7.6$ GeV exists practically unmodified (see the upper panel of Fig.\ 2 in \cite{Bugaev:2016ujp}).
At the same time the regime of  reaching the strangeness chemical equilibrium, i.e.    $\gamma_s =1$, at  $\sqrt{s_{NN}} \ge  8.8$ GeV is also observed in \cite{Bugaev:2016ujp}. 

Based on this discussion we can further  refine a hypothesis on the cause of chemical equilibrium of strange charge  in central nuclear collisions  at the collision 
energy $\sqrt{s_{NN}} \ge  8.8$ GeV formulated in  \cite{Sagun17}. In our opinion,   the most natural explanation of 
this phenomenon  is that at  $\sqrt{s_{NN}} \ge  8.8$ GeV there appeared the quark-gluon bags with the exponential mass spectrum 
proposed by R. Hagedorn \cite{Hagedorn}. As it was predicted in \cite{Thermostat1} and shown numerically in \cite{Hstate1,Hstate2,Hstate3}, the exponential mass spectrum acts as a perfect thermostat and a perfect particle reservoir.
In other words, all particles which appear from such bags at their hadronization will be born in a state of  full thermal and chemical equilibrium 
\cite{Hagedorn,Hstate1,Hstate2,Hstate3}.  

Furthermore,  an existence of two peaks of $\gamma_s$ strongly enhances 
the hypothesis about existence of two phase transformations.
Indeed, since the right slope
of the  higher peak contains the region  of  the 1-st order phase transition at $\sqrt{s_{NN}} =  4.3-4.9$ GeV, then 
it is naturally to assume that the right  slope
of the lower  peak also contains the phase transition region. 
Moreover,  from Fig.\ \ref{Fig4} one can see 
that the end of the 1-st order phase transition is shifted on  1.1 GeV to a higher collision energy than the peak existing at   $\sqrt{s_{NN}} =  3.8$ GeV.  Then in accordance with our hypothesis the end of the other phase transition  should be  also
shifted on about 1.1 GeV to a higher  collision energy  than the peak at $\sqrt{s_{NN}} = 7.6$ GeV,  i.e. one would expect it at 
$\sqrt{s_{NN}} \simeq 8.7$ GeV. Thus, again we independently arrive to the same conclusion as from the analysis of the 
trace anomaly and the baryonic density peaks that the vicinity of  $\sqrt{s_{NN}} \simeq 8.8-9.2$ GeV  corresponds to 
a phase transformation.  Since this conclusion is based on the assumption that the 1-st order phase transition can be reached at 
$\sqrt{s_{NN}} = 4.3-4.9$ GeV, then it is logically to expect that QCD has a 3CEP and it is located close to 
the collision energy range  $\sqrt{s_{NN}} \simeq 8.8-9.2$ GeV. Thus, quite independently to Ref.  \cite{Ko2017} we came to a similar conclusion about the endpoint, but, in contrast to   \cite{Ko2017}, we consider it as the 3CEP. 

Note that the collision energy  $\sqrt{s_{NN}} \simeq 8.8$ GeV as  the onset of deconfined phase appears not only  in Ref.
\cite{Ko2017}.  Thus,   in Ref.  \cite{Nayak2010} it was shown that the horn in  ${K^+}/{\pi^+}$  ratio can be naturally
explained, if one assumes that the partonic phase, or the   quark gluon plasma (QGP)  production threshold  is   
$\sqrt{s_{NN}} \simeq 8.8$ GeV.  Note that we came to a similar conclusion  above on the basis of the Hagedorn thermostat hypothesis. 
But now we are facing a question, if  the QGP is formed in heavy ion collisions at  $\sqrt{s_{NN}} \simeq 8.8$ GeV,
then what kind  of phase transition occurs at lower collision energies  $\sqrt{s_{NN}} = 4.3-4.9$ GeV? 
From the discussion above it is evident that such a phase cannot consist from the bags of QGP which are filled 
with the quarks and gluons, otherwise one would find $\gamma_s=1$.

\label{sec:strwqlbr}
\section*{Strangeness equilibrium dale and mixed phase as an explicit thermostat}

Before discussing the possible interpretations of the above results, we would like to reinforce our arguments  about the mixed phase formation at  $\sqrt{s_{NN}} = 4.3-4.9$ GeV.
Using  the thermostatic properties of the mixed phase of the 1-st order phase transition \cite{Thermostat1} below 
we would like to explain the appearance of  the strangeness equilibrium dale, i.e. $\gamma_s \simeq 1$,  at $\sqrt{s_{NN}} = 4.9$ GeV.
 In 
\cite{Thermostat1} the examples of the explicit thermostat and explicit particle reservoir were discussed and it was argued that under the constant pressure condition  the mixed phase  of the  1-st order phase transition, i.e.  two  pure phases being in a  thermal and chemical equilibrium with each other, 
represent both thermostat and particle reservoir as long as  it has enough energy resources to keep a constant temperature.  In  other words, under the constant pressure condition an explicit thermostat keeps a constant temperature despite transmitting out (in) some amount of heat. The latter only changes the volume fractions of two phases: the phase with higher heat capacity (for definiteness,   a liquid)  condenses a certain amount of gas  under external cooling  or    it partly evaporates  into a gas under external  heating.  Apparently,  for finite systems the amount of imparted  heat  is also finite and depends on the masses of both phases and their heat capacities.  Similarly, one can add or remove some amount of 
each phase, but  under the constant pressure condition the system will continue to  keep a constant temperature and a full chemical equilibrium, i.e. the equal values of chemical potentials for both pure phases.
For our discussion it is important that  up to some maximal   value  any amount of  the removed phase, namely the gas of hadrons, for definiteness,  will be,  by definition,  in a full chemical and thermal equilibrium with the mixed phase. Again the maximal amount of removed phase depends on  the masses of both phases and their  energy  densities.  Note that to justify  such a picture 
we have to assume that the mixed phase has sufficiently large volume, so 
the finite size effects are not significant. 

\begin{figure}[th]
\centerline{\hspace*{-0.44mm}
\includegraphics[width=84mm]{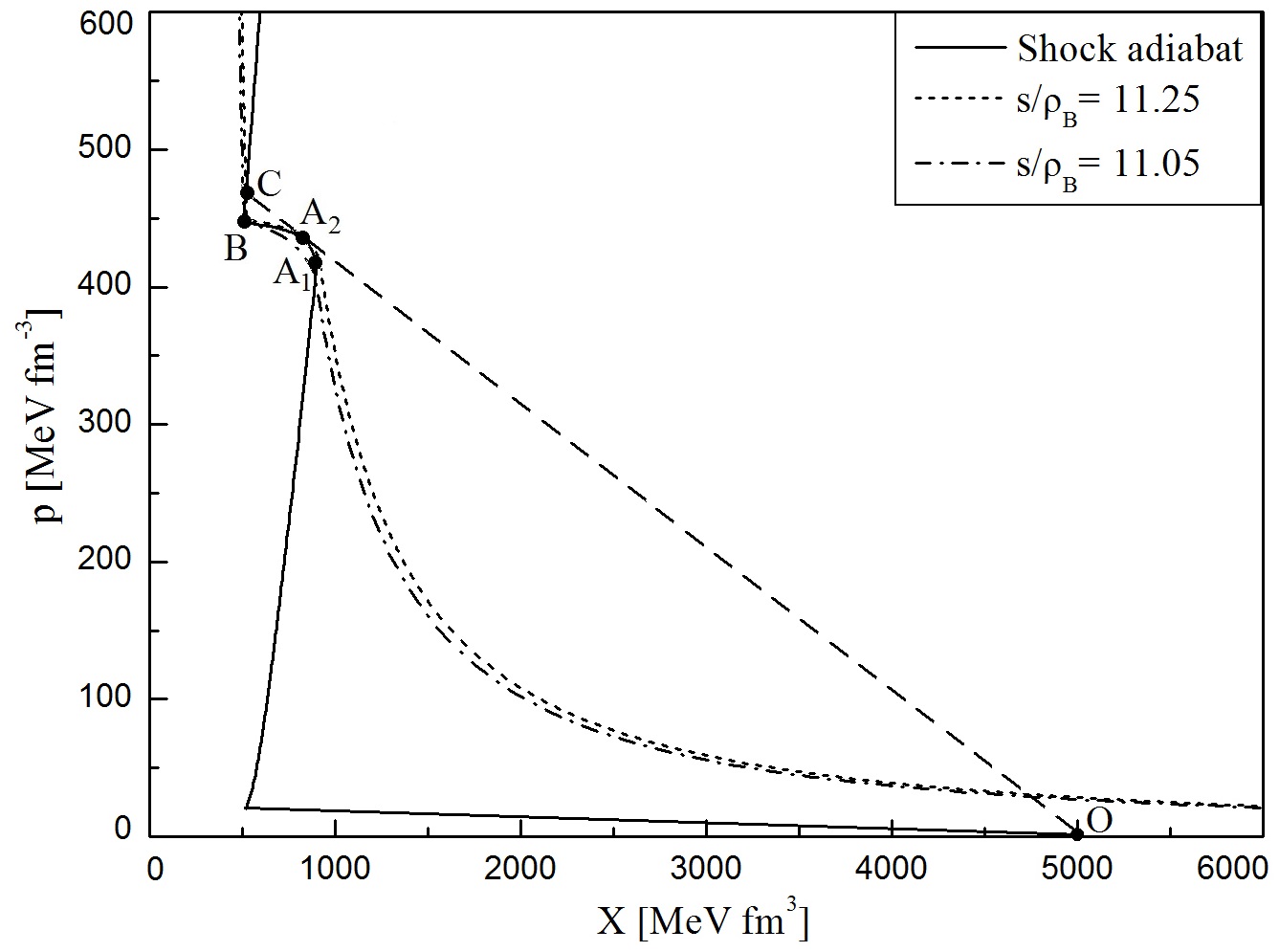}
\hspace*{2.2mm}
\includegraphics[width=84.mm]{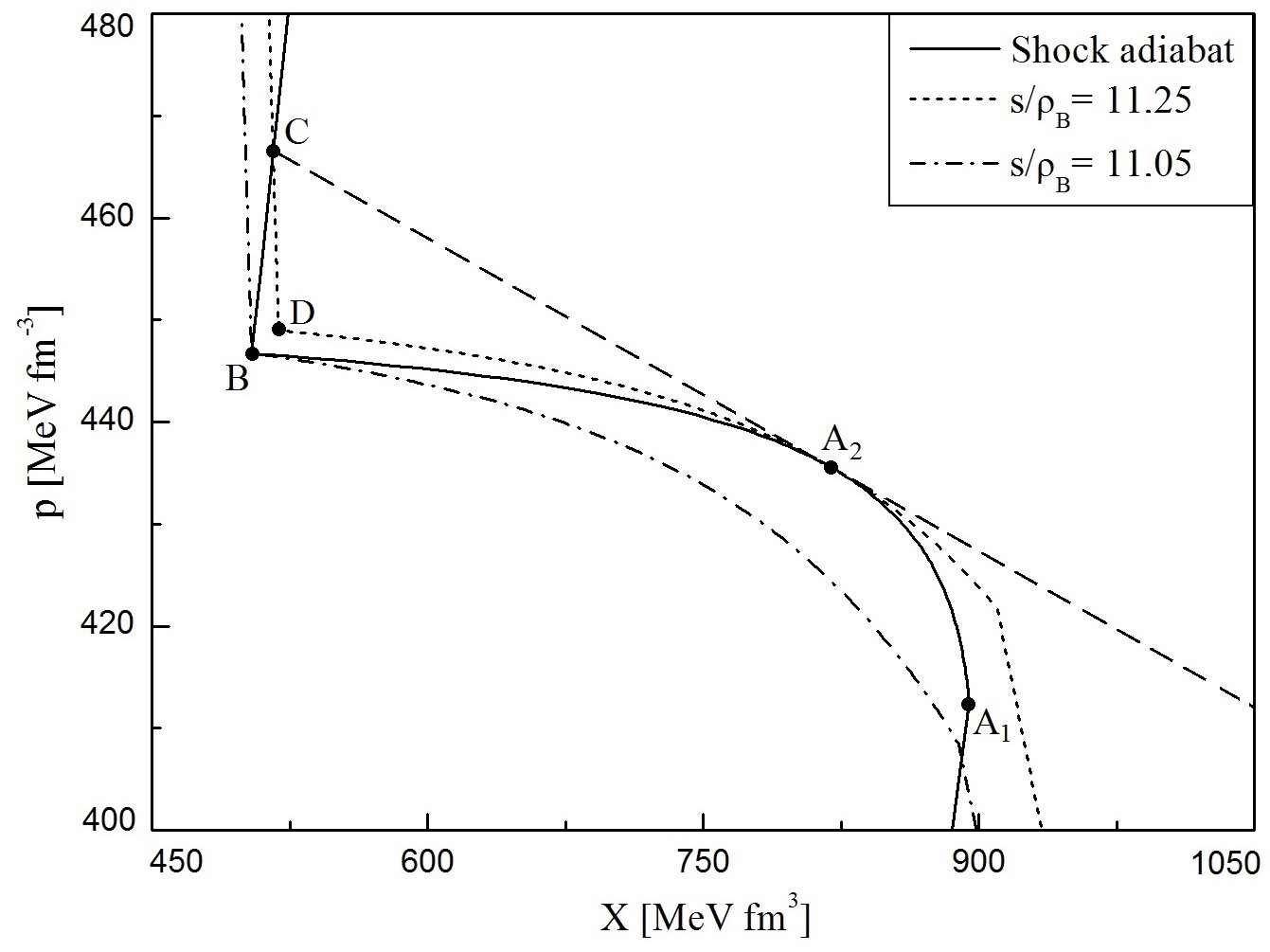}
 } 
 \caption{ The compressional  shock adiabat OA$_1$A$_2$BC (solid curve)  in the $X-p$ 
plane found in \cite{Bugaev:2014,Bugaev:2015}. It is calculated for the EoS with first-order phase transition discussed in the text.  
The segments OA$_1$, A$_1$B, and BC of the adiabat 
correspond to the hadronic, mixed and  new  phases (see below for its interpretation), respectively.   
Shock transitions into the region of states A$_2$BC are mechanically unstable. 
Hence  the segment A$_2$BC should be replaced by the generalized shock adiabat taken from \cite{Bugaev:2014,Bugaev:2015}.
The tangent 
point A$_2$ to the shock adiabat  (see the dashed line) is the Chapman-Jouguet 
point \cite{Landavshitz}.
The dotted and dash-dotted curves show the Poisson adiabats with 
values of entropy per baryon specified in the legend. The right panel shows more details  inside the mixed phase region.
 For more explanations see the text.}
  \label{Fig5}
\end{figure}

Above we discussed a static picture, but the question is whether the condition of constant pressure can be achieved in central nuclear   collisions  at   $\sqrt{s_{NN}} = 4.3-4.9$ GeV.  
To extend this picture  to the  case of central collisions  of heavy ions  we  refer  to  the compressional 
shock model of these collisions 
\cite{Mish:78,Horst:80,Kamp:83,Horst_PR:86,Barz:85,Satarov:11}
and its  generalization 
to the mixed phase with the anomalous thermodynamic properties \cite{GSA1,GSA2,Bugaev:2014,Bugaev:2015}. 
This model of the  collision process, which neglects the nuclear transparency, can be reasonably well  
justified at intermediate collision energies per nucleon $1$ A GeV $ \le E_{lab} \le $ 20 A GeV \cite{Horst:80,Kamp:83,Horst_PR:86,Barz:85,Satarov:11}. 
Moreover, the direct comparison  of the results of the compressional shock model 
with those of the three-fluid model \cite{3fluid}
made  in Ref.\ \cite{Satarov:11} demonstrates once more that
at laboratory energies up to 30 A GeV this model can  
be used for quantitative estimates, while at higher energies (up to about 50 A GeV in laboratory frame) it provides a 
qualitative description only.

\begin{figure}[th]
\centerline{\hspace*{-0.44mm}
\includegraphics[width=84mm]{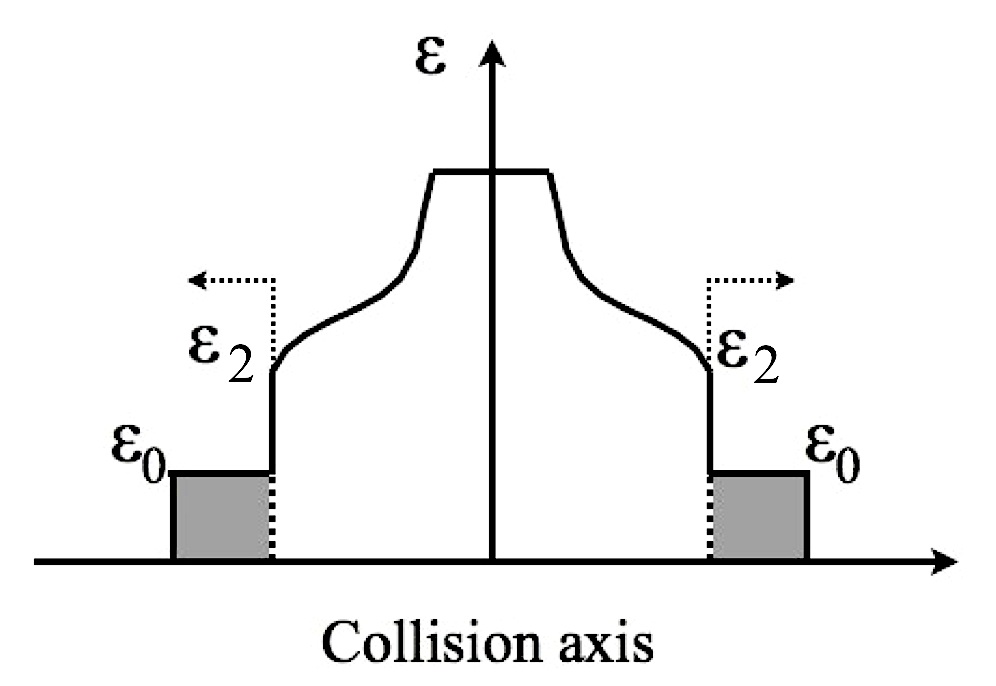}
 } 
 \caption{Sketch of the energy density profile for a central collision of two nuclei (grey 
 areas) that corresponds to a stable flow pattern  to the mixed phase region A$_2$D of Fig.\ \ref{Fig5}.
 The dashed arrows show the 
direction of  shock propagation in the center of mass frame. Two shocks between the states $\varepsilon_0$ (the point O in Fig.\ \ref{Fig5}) and $\varepsilon_{2}$ (the point A$_2$ in Fig.\ \ref{Fig5})  are followed by the compressional simple waves. For more details see the text.
   }
  \label{Fig6}
\end{figure}

The compressional shock model of central nuclear collisions 
 allows one to determine the  initial conditions for the subsequent hydrodynamic evolution. The latter are given 
 by the  shock adiabat \cite{Landavshitz}, each point of which corresponds to a certain energy of collision (more details can be found in \cite{Satarov:11,GSA1}). 
The shock adiabat   OA$_2$BC  (see Fig.\ \ref{Fig5}) which was found in \cite{Bugaev:2014,Bugaev:2015} shows  that in the mixed phase (the segment 
A$_1$A$_2$B in Fig.\ \ref{Fig5}) of the 1-st order phase transition the pressure at the central longitudinal rapidity region  practically does not depend  on the collision energy.  Note, however, that the shock transitions into the states belonging to the segment A$_2$BC for the shock adiabat   OA$_2$BC   of   Fig.\ \ref{Fig5}  are mechanically unstable \cite{Zeldovich,Rozhd} due to anomalous thermodynamic properties  of the mixed phase (the segment A$_2$B) and in this case  more complicated flow patters should appear. The latter for the relativistic shock waves were  determined in \cite{GSA1,GSA2} and were calculated in  \cite{Bugaev:2014,Bugaev:2015}. In particular, the shock adiabat region of mixed phase  A$_2$B should be replaced by the the Poisson  adiabat   \cite{Landavshitz}   A$_2$D (see the right panel of  Fig.\ \ref{Fig5}; here and below the variable $X \equiv \frac{(\epsilon + p)}{\rho_B^2}$ denotes  the generalized specific volume \cite{Landavshitz}).  
It is remarkable that from the point A$_2$ to  the point D on the Poisson adiabat the pressure changes from about 440 MeV$\cdot$fm$^{-3}$ to about  450 MeV$\cdot$fm$^{-3}$, i.e. the relative change of pressure is about $1/44$ or 2.27\%. 
The flow configuration which corresponds to the segment A$_2$D  consists of the compressional  shock wave and compressional  simple wave moving outwards (see Fig.\ \ref{Fig6}).
The shock wave describes the transition 
 from the cold nuclear matter state $(X_0; p_0 =0 )$ (the point O in Fig.\ \ref{Fig5}) to the state $(X_{A_2}; p_{A_2} )$, i.e. the tangent point A$_2$ to  the shock adiabat,   while the compressional  simple wave  describes the transformation  from the state  $(X_{A_2}; p_{A_2} )$  to any state $(X; p)$ on the considered segment of the Poisson adiabat with  $X_{A_2} \ge X \ge X_D$ and $ p_{A_2} \le  p \le p_D$  (see Refs. \cite{GSA1,GSA2,Bugaev:2014,Bugaev:2015} for more details).  
In other words,  by increasing the collision energy, 
at the central rapidity region of the collision 
one creates the states with higher energy density, but,  practically,  with the constant pressure.   
Also due to the fact that the energy density 
of  created dense  phase (the phase of massless particles or (PMP) hereafter) is about 10 times higher than the one of the hadron gas \cite{Bugaev:2014,Bugaev:2015}, the major  part of energy of  a  matter  formed after the compressional  shocks between the states $(X_0; p_0=0 )$   and $(X_{A_2}; p_{A_2} )$ disappear   is concentrated right at the central  rapidity region of collision.  Therefore, just the mixed phase states with practically the same pressure and the same value of entropy per baryon in each 3-dimensional point of the formed matter should define the pattern of its hydrodynamic  expansion.  Apparently, the hadronic matter  which will  be born from the mixed phase will be in a full thermal and chemical equilibrium with it.  Its further hydrodynamic evolution is out of the scope of the present work, but 
according to the contemporary paradigm the hadronic matter, including the strange hadrons,  born off  the mixed phase  will remain in the full equilibrium till the moment of  CFO. 

If the collision energy increases above the point D on the Poisson adiabat, then the matter formed at the central  rapidity region of collision  will correspond to the segment $DC$ and the part of mixed phase will gradually  decrease vanishing completely at the point C of the shock adiabat.  Simultaneously, the role of  mixed phase  as the explicit thermostat  and particle reservoir will be 
gradually diminished.  Counting the discussion above, we can explain the $\gamma_s=1$ dale seen  in the right panel of Fig.\ \ref{Fig4} by   the mixed phase formation at the  collision energies   $\sqrt{s_{NN}} = 4.3-4.9$ GeV and   
its   steady disappearance  at  $\sqrt{s_{NN}} = 6.3-7.6$ GeV together with the corresponding appearance of chemical nonequilibrium of strangeness.
Furthermore, such  a hypothesis allows  one to  simultaneously understand  the strong increase of the modified Wroblewski factor $\lambda_s$  within  the 
collision energy interval $\sqrt{s_{NN}} = 2.7-7.6$ GeV and  the nontrivial   collision energy dependence 
of   the $\gamma_s$ factor at these energies (compare the both panels  of  Fig.\ \ref{Fig4}).

Although it is expected that  in QCD there may exist two phase transitions, i.e. deconfinement and chiral symmetry restoration,  it is hard to unambiguously  identify the nature of  phase transition at $\sqrt{s_{NN}} = 4.3-4.9$ GeV, since  presently 
there is no reliable information from lattice QCD at high baryonic densities.   In Refs. \cite{Bugaev:2014, Bugaev:2015,Bugaev:2016a} it was concluded that at  these collision energies there may exists the QGP which is probed by the lattice QCD approach.
According to our present hypothesis in heavy ion collisions the  two phase transitions may be observed.  From the discussion above 
it is clear that  at the collision energies   $\sqrt{s_{NN}} = 8.8-9.2$ GeV there may exist transition to the QGP, i.e. the phase of matter studied by the lattice QCD, and, hence, at the collision energies   $\sqrt{s_{NN}} = 4.3-4.9.$ GeV the transition to QGP is impossible. Thus, the  latter  collision energy range  may correspond to another phase which we call 
the phase of massless particles (PMP).  
Although at present the   amount of experimental data at the collision energy range $\sqrt{s_{NN}} = 4.9-9.2$ GeV is rather limited, 
we can make an educated guess about its properties. Our  guess is based on the  EoS  of  the PMP  which 
 was determined in Refs. \cite{Bugaev:2014,Bugaev:2015}  from fitting the entropy per baryon  $s/\rho_B$ along the  shock adiabat \cite{GSA1,GSA2}. This  EoS is similar to the MIT-Bag model \cite{MITBag}
\begin{eqnarray}\label{EqIX}
p_{PMP} =A_0T^4+A_2T^2\mu^2+A_4\mu^4-B  \,, 
\end{eqnarray}
but the constants  $A_0 \simeq 2.53 \cdot 10^{-5}~{\rm MeV}^{-3}{\rm fm}^{-3}$, $A_2 
\simeq 1.51 \cdot 10^{-6} ~{\rm MeV}^{-3}{\rm fm}^{-3}$, $A_4 \simeq 1.001 \cdot 10^{-9}~{\rm 
MeV}^{-3}{\rm fm}^{-3}$, and $B \simeq 9488~{\rm MeV}~{\rm fm}^{-3}$ are rather different 
from what is predicted by the perturbative QCD for massless gluons and  (anti)quarks.  Despite this fact,  an additional condition that
the pseudocritical temperature value  at zero baryonic density is  about 150 MeV
was used in the fitting procedure of Ref.  \cite{Bugaev:2014}. Note that the  
value $T = 150$~MeV  is 
 in agreement with the lattice QCD data \cite{Aoki}.  

In  Ref.  \cite{Bugaev:2014}  the difference of    coefficients $A_0$, $A_2$ and $A_4$
from the ones of perturbative QCD $A_0^{pQCD}, A_2^{pQCD}$, and $A_4^{pQCD}$  was interpreted
as the  $T$ and $\mu_B$ dependence of  non-perturbative  pressure $B_{eff}(T, \mu_B) \equiv B - (A_0-A_0^{pQCD} )T^4 - (A_2-A_2^{pQCD})T^2\mu^2 - (A_4 -A_4^{pQCD})\mu^4$ at high baryonic densities.  Then the pressure  (\ref{EqIX})
can be rewritten as  $p  =A_0^{pQCD}T^4+A_2^{pQCD}T^2\mu^2+A_4^{pQCD}\mu^4-B_{eff} $. 
This was a way  to avoid the contradiction with perturbative QCD. 
However,  now, in accordance with the hypothesis of two phase transitions,    one can 
consider  Eq. (\ref{EqIX})  differently, namely  as a source of  information about the PMP  which is formed at the collision
energies 4.9 GeV $ < \sqrt{s_{NN}} < 8.8$ GeV. 

Recall that the EoS for massless bosons and  fermions (with the  chemical potential $\mu$)  has the form \cite{Chin78}
\begin{eqnarray}\label{EqX}
p_{massless~ gas} = \frac{\pi^2 N_{dof}^{eff}}{90} \, T^4+ \frac{N_{f}^{eff}}{12}\,T^2\mu^2+ \frac{N_{f}^{eff}}{24\,  \pi^2}\,\mu^4  \,, 
\end{eqnarray}
where the  number  of the massless degrees of  freedom 
$N_{dof}^{eff} = N_{bos}^{eff} + \frac{7}{4} N_{f}^{eff}$
is given in terms of degeneracy factors of  massless  bosons $N_{bos}^{eff}$ and 
massless (anti)fermions  $N_{f}^{eff}$.
Comparing the  coefficients of powers of temperature in Eqs. (\ref{EqIX}) and (\ref{EqX}) one  can estimate  the number of the massless degrees of  freedom of this phase $N_{dof}^{eff}$ from the coefficient $A_0$.
Then  one obtains  the number of massless boson and fermion degrees of freedom  $N_{dof}^{eff} =  \frac{90}{\pi^2} \, A_0 \hbar^3  \simeq 1770$, where we used  the Planck constant $\hbar$ in order to get a dimensionless value of  $A_0$. Note that $N_{dof}^{eff}$  is essentially larger than the number of massless degrees of freedom
of QCD (the total degeneracy of gluons, quarks and antiquarks), but it is comparable to the total number of  spin-isospin configurations of known hadronic states \cite{PDG}. 
Assuming that fermions have the baryonic charge $\pm 1$,  one can use   the coefficient $A_2$ to estimate the number of  massless fermionic and antifermionic degrees of freedom  $N_{f}^{eff} =  12 \, A_2 \hbar^3  \simeq  141$, then one finds that  it  is still far larger than the expected number of  quarks and antiquarks.   Also one can use the coefficient $A_4$ to get  the number of  massless fermionic and antifermionic degrees of freedom  $\widetilde N_{f}^{eff} =  24 \pi^2 \, A_4 \hbar^3  \simeq  1.82$ which is inconsistent with $N_{f} ^{eff}  \simeq  141$ found from the coefficient $A_2$. Clearly, it would be a great surprise, if  the coefficients  $A_2$ and  $A_4$ of Eq.   (\ref{EqIX})  would completely agree with the corresponding coefficients in Eq.  (\ref{EqX}). Hence, we conclude that interaction between  the constituents of PMP  may differ from the simple bag model parameterization.  
Nevertheless, our hypothesis  is  that   the  PMP  consist of  non-strange hadronic states with the  mass $m_{eff}$ which is much smaller  than $T$, i.e.   $m_{eff} \ll T$.  In this case  the PMP can be considered as the hadronic state with the restored  unitary symmetry. 
{Our hypothesis on the PMP existence   is strongly supported by the results of the microscopic model Parton-Hadron-String Dynamics  (PHSD)  \cite {Cassing16a,Cassing16b} which show that at the vicinity of the collision energy $\sqrt{s_{NN}} \simeq 7$ GeV the effects of  baryonic and partonic mass reduction, i.e. the effects of chiral
symmetry restoration in the hadronic phase, are very strong.}

The question about existence of   almost  massless strange hadrons in the PMP  cannot be answered at the moment with high confidence,
i.e. now it is unclear  whether the unitary symmetry in PMP  is restored completely or in the non-strange sector only.   However, looking at the $\gamma_s$ peaks at  $\sqrt{s_{NN}} = 3.8$ GeV and  $\sqrt{s_{NN}} = 7.6$ GeV in the right panel of  Fig.\ \ref{Fig4} one may guess that  the same mechanism of the strangeness enhancement may be responsible for them, while this mechanism disappears at  $\sqrt{s_{NN}} \ge 8.8$ GeV due to the 
formation of  quark-gluon bags with the  Hagedorn mass spectrum. Therefore, if  in the hadronic phase, i.e. at  $\sqrt{s_{NN}} < 4.3$ GeV the strange hadrons are massive, then we believe  it is reasonable to assume that in the PMP  they are massive too. 
{Although the  PHSD microscopic model predicts an essentially weaker  reduction of the constituent strange quark 
compared to the mass of  constituent u- and d-quarks \cite {Cassing16b}, we believe that
the questions of whether and how  the masses of strange hadrons  in the PMP   are reduced compared to the vacuum values,  
will have to be answered experimentally on the  accelerators   of new generation, i.e. NICA JINR and FAIR GSI.
}
 
It is interesting that a  long time ago an existence of  the hadronic phase with abundant number of baryon-antibaryon pairs 
was discussed in Ref.  \cite{Glend87}. In this model the  non-linear interaction of  scalar  meson   provides an essential reduction 
of baryonic masses and generates an additional solution to  the usual hadronic matter which in Ref.  \cite{Glend87} is called as the
abnormal matter.  Similar  additional solutions for the self-interacting mesonic fields  are also discussed in \cite{Glend79,Theis83,Takagi84}.   
It is intriguing, however,  that some thermodynamic parameters of the normal/abnormal matter are close to our findings. For example, on the isotherm $T=148$ MeV shown in Fig.\ 3 of   \cite{Glend87} the maximal pressure of the usual hadronic matter
(normal state according to Ref. \cite{Glend87}) is about 420 MeV$\cdot$fm$^{-3}$. Just compare it with the pressure of point A$_1$ on the boundary of hadronic and mixed phases in Fig.\  \ref{Fig5}.  On the other hand,  from Fig.\ 3 of   \cite{Glend87} one can see that along the isotherm $T=148$ MeV the abnormal  matter  can acquire 
higher values of  pressure which are comparable to  the ones shown in Fig.\  \ref{Fig5} of this work  for the compressional and generalized  shock adiabats.

 Of course, the existence of   PMP  with nearly massless non-strange hadrons  is not the only possibility.  Another possibility could be   a presence of  tetraquark condensate, which may generate two 1-st order chiral phase transitions  \cite{TetraQ}. One more  possibility could be  an existence of  the quarkyonic matter   \cite{QYON}, which has an appropriate collision energy range.  Hopefully, the nature and properties of  the PMP  will   be studied at the accelerators  NICA JINR and FAIR GSI.
 However, despite of its origin, if at high temperatures one takes the EoS (\ref{EqIX}) literally, then one apparently  faces a contradiction with  the lattice QCD   thermodynamics.  
Indeed, if one fixes $\mu$ and increases $T$, then despite of the large value of constant pressure $B$ in Eq. (\ref{EqIX}) at   high temperatures the  pressure of PMP $p_{PMP}$   will exceed the pressure of massless  (anti)quarks and gluons.
The first possibility to avoid such a contradiction is to claim that the PMP  pressure (\ref{EqIX}) can be used just in the 
vicinity of the phase transition, since it does not correctly account for interaction of constituents at higher energy density. 
Basically we agree with this statement, but, in addition, we  consider a possibility to get rid of the discussed contradiction with lattice QCD on  a physical ground. 

\label{sec:repuls}
\section*{Phase of massless particles  with relativistic hard-core repulsion}

If, according to our hypothesis,  the PMP  consists of hadrons, then one cannot ignore their hard-core repulsion.  Nowadays  there are many articles discussing various applications of  the  hard-core repulsion in high energy nuclear physics   \cite{SFO,Veta14,Bugaev:2016,Sagun17,Bugaev17,VDW1,HRGM:13,Lattimer:1991nc,Shen:1998gq,Typel16,RelVDW1,RelVDW2},  however, the rigorous approach  to relativistic treatment of  Lorentz contraction of  rigid spheres was developed in Refs.  \cite{RelVDW1,RelVDW2}.  The ultra-relativistic limit of Lorentz contracted  rigid spheres   analyzed  in  \cite{RelVDW2} shows that in contrast to the non-relativistic  case   the EoS  with  relativistic Van der Waals hard-core repulsion  is causal at high pressures, i.e. its speed of sound does not exceed the speed of light. 

Our suggestion  to avoid  a contradiction with the lattice QCD thermodynamics is to account for the Lorentz contraction of  rigid spheres in the PMP  pressure (\ref{EqIX}) in the spirit of Ref. \cite{RelVDW2}. For simplicity we consider the  Boltzmann limit of a one component gas, i.e.  for  equal masses of bosons and (anti)fermions and the same hard-core radius $R_0$. Note  that such a  simplification  is in line with  the main property  of  PMP  that particle masses are negligible compared to system temperature  and simultaneously it   provides  us with a qualitatively correct solution of this  complicated task.  The pressure of the Lorentz contracted rigid spheres $p_{rel}   (T, \mu) $   having the temperature $T$,  the chemical potential $\mu$ and the  relativistic excluded  volume per particle $v (\1 k_1,\1 k_2)$ is given by a solution of the system  \cite{RelVDW2}
\begin{eqnarray}  \label{EqXI}
p _{rel}  (T, \mu) 
& = & g \, F_{rel}  (T, \mu, p_{rel} ) \,,
 \\
 \label{EqXII}
F_{rel}  (T, \mu, p )  &\equiv & \frac{T }{ \rho_{th} (T)} {\textstyle  \exp \left[ \frac{\mu}{T}\right]}
\int 
\frac{d{\1 k_1}}{(2\pi)^3}
\frac{d{\1 k_2}}{(2\pi)^3}
\, \, {\textstyle \exp\left[- \, \frac{  v (\1 k_1,\1 k_2)\, p   \, +\,E(k_1)\, + \,E(k_2) }{T} \right] }
\,,~~ \\
 \rho_{th} (T) & =& \int \frac{ d{\1 k} }{(2\pi)^3} {\textstyle  \exp\left[- \,  \frac{ \,E(k)\,  }{T} \right] } \,\,,
 \label{EqXIII}
\end{eqnarray} 
where
$m$ is  the mass of hadron,  $g$ is its degeneracy factor, 
 $ \rho_{th} (T)$ denotes the thermal density of particles per single degree of freedom,  $E( k) = \sqrt{k^2 + m^2}$
is relativistic energy of particle having the 3-momentum $\vec k$ and $\mu$ is its  baryonic chemical potential.  In the ultra-relativistic 
limit the relativistic excluded  volume per particle $v (\1 k_1,\1 k_2)$ can be cast as \cite{RelVDW2}
\begin{equation} \label{EqXIV}
v(\1 k_1, \1 k_{2})  \approx   \frac{  v_0 }{2 }  
\left[  \frac{m}{E(\1 k_1)} + \frac{m}{E(\1 k_2)} \right]
\left[ 1 + \cos^2 \left(  \frac{\Theta_v}{2} \right) \right]^2  
 +  \frac{3 \, v_0 }{2}  \sin \left( \Theta_v \right)   \,\,,
\end{equation} 
where the non-relativistic proper volume of particle with the hard-core radius $R_0$  is $v_0 = \frac{4}{3} \pi R_0^3$.
Eq. (\ref{EqXIII}) is valid for $0 \le \Theta_v \le \frac{\pi}{2} $; to use it for 
$ \frac{\pi}{2} \le \Theta_v \le \pi $ one has to make a replacement 
$ \Theta_v \longrightarrow \pi -  \Theta_v$ in  (\ref{EqXIII}). 
The coordinate system   is chosen in such a way that the angle $\Theta_v$ between the 3-vectors of 
particles' momenta $\1 k_1$ and  $\1 k_{2}$ coincides with the usual spherical angle $\Theta$ of  spherical
coordinates (for more details see Appendix A in \cite{RelVDW2}). 
For definiteness, the OZ-axis of the momentum space  coordinates of  the second particle  is chosen to coincide with the 3-vector of the momentum $\1 k_1$ of the first particle.

It is evident that the  volume  $v (\1 k_1,\1 k_2)$ strongly depends not only on particle momenta, but  on  the angle between them. 
Due to this property at high pressure $p_{rel}$ the contribution of configurations with larger volume $v (\1 k_1,\1 k_2)$ are suppressed and, hence, the main contribution to the double integral in  Eq. (\ref{EqXI}) is given by the configurations with the minimal value of  relativistic excluded  volume. It is remarkable that with an accuracy of about 7\% the ultra-relativistic expression
(\ref{EqXIV}) recovers the excluded volume of two non-relativistic rigid spheres  \cite{RelVDW2} and, therefore, it can be  safely used at  low particle densities too.  

Using the   pressure  of Lorentz contracted rigid spheres (\ref{EqXI}), for $m \ll T$, one can generalize  EoS  (\ref{EqIX}) to high 
energy densities as
\begin{eqnarray}\label{EqXV}
\tilde p_{PMP}  &\simeq &   T \pi^2 (A_0 - 2A_2) F_{rel}  (T, 0, \tilde p_{PMP})  +  T \pi^2 A_2  [ F_{rel}  (T, \mu, \tilde p_{PMP})  \nonumber \\
&+& 
 F_{rel}  (T, -\mu, \tilde p_{PMP})  ]  
-  B - \left( \frac{A_2}{12} - A_4 \right) \mu^4 \,.
\end{eqnarray}
The first term on the right hand side of  Eq.  (\ref{EqXV})  corresponds to  the Boltzmann  gas of chargeless  particles,
while the second and third ones correspond to  the Boltzmann  gas of  particles with charges +1 and -1,  respectively.
Expanding the exponents $\exp (\pm \mu/T) $ which are staying  on the right hand side of  (\ref{EqXV})  up to terms $(\mu/T)^6$, one can readily check that for $m \ll T$ and $\frac{v_0\, \tilde p_{PMP}}{T} \ll 1$ the  right hand side of    (\ref{EqXV}) 
recovers the one of  Eq.  (\ref{EqIX}).    The  term  $\frac{A_2\, \mu^4}{12}$ on the right hand side of   (\ref{EqXV})  
``compensates"  an extra contribution coming from the expansion of   fermionic  exponents $\exp (\pm \mu/T) $. 
It is easy to verify that for massless particles and a zero value of hard-core radius $R_0$ the $T$ and $\mu$ dependent terms 
of pressures  (\ref{EqIX}) and  (\ref{EqXV})  differ from each other by less than 10\% for  $\left| \frac{\mu}{T} \right| \le 2.8$ and any $T$.

Apparently,  Eq. (\ref{EqXV})    is not a unique  way to extrapolate  the PMP pressure (\ref{EqIX})  to high energy densities, but,
besides the fact that the pressure (\ref{EqXV})  is one of the simplest possibilities, one can also check that for $T\ge 137.1$ MeV 
and $m \ll T$ it has only a single positive  solution $\tilde p_{PMP} (T, \mu) > 0$ for any value of  $\displaystyle \frac{\mu}{T}$.  Therefore,  we believe that  Eq.  (\ref{EqXV}) is well suited  for  demonstrating   how the Lorentz contracted  relativistic excluded volume can  modify     the  PMP  pressure at high energy densities.  A sufficient condition 
for an existence of a single positive  solution $\tilde p_{PMP} (T, \mu)$ of (\ref{EqXV}) can be found as follows. For  the fixed values of $\mu$ and $T$ with the restriction  $m \ll T$   one can solve Eq.  (\ref{EqXV})  graphically. Noting that the right hand side 
of  (\ref{EqXV})  is a monotonously decreasing function of  the variable  $\tilde p_{PMP}$  due to excluded volume effect, one can  conclude that  such a function can have a single intersection with the straight line $p = \tilde p_{PMP}$, i.e. with the left hand side of  (\ref{EqXV}), if   at the point $p = 0$ the inequality 
\begin{eqnarray}\label{EqXVI}
 && T \pi^2 (A_0 - 2A_2) F_{rel}  (T, 0, 0)  +  T \pi^2 A_2  \left[ F_{rel}  (T, \mu, 0) + F_{rel}  (T, -\mu, 0)  \right]   \nonumber \\ 
&& - B - \left( \frac{A_2}{12} - A_4 \right) \mu^4 > 0 \,,
\end{eqnarray}
is obeyed. 

Assuming that the condition (\ref{EqXVI}) is obeyed,  we consider the asymptotic behavior of the pressure (\ref{EqXV}) for two cases, namely high $T$ limit and high $\mu$ limit.
Apparently, both of them correspond to a high pressure limit and, therefore, the bag pressure $B$ and the last term on the right hand side of  Eq. (\ref{EqXV}) can be neglected.  
Applying  the  results of Ref.  \cite{RelVDW2} to  Eqs.  (\ref{EqXII}), one can easily show that in the limit  $T \rightarrow \infty$  the dimensionless  variable   $z \equiv  \frac{2 m \, v_0 \, p_{rel} }{T^2} \rightarrow Const < \infty$. Using this  result for the pressure  (\ref{EqXV}),   in the limit  $T \rightarrow \infty$  and finite values of $\mu$ one can find 
\begin{eqnarray}\label{EqXVII}
\tilde p_{PMP}  (T, \mu) \rightarrow   \left[ \frac{\pi^2 A_0 + 2 \pi^2 A_2 ( \cosh\left[ \frac{\mu}{T} \right] -1) }{v_0^2}  \right]^\frac{1}{3} T^2 \, f (m) \,, 
\end{eqnarray}
where   $ f (m)  \sim O(1)$ is a slowly decreasing  dimensionless function  of hadronic   mass $m$. The effective number of 
degrees of freedom now is $ [\pi^2 A_0 \hbar^3]^\frac{1}{3} \simeq  12.4$. 
Moreover, the temperature dependence of pressure  (\ref{EqXVII}) is essentially weaker than the one for massless quarks and gluons (see Eq. (\ref{EqX})).  

In the limit of fixed $T$ and   $\mu \rightarrow \infty$ one finds for Eqs.  (\ref{EqXI}) and, similarly,  for  (\ref{EqXV}) that  \cite{RelVDW2}
\begin{eqnarray}\label{EqXVIII}
\tilde p_{PMP}  (T, \mu) \rightarrow   \frac{\mu^2}{32 \, m\, v_0}  \ \,, 
\end{eqnarray}
the pressure of relativistic hard-spheres does not depend on the number of degrees of freedom at all and its $\mu$ dependence 
is again essentially weaker than $\mu^4$ dependence of massless (anti)quarks. Therefore,  these two examples  show that the Lorentz contraction of relativistic 
hard-spheres, in principle,  is able to resolve the problem of  high energy density  extrapolation  of  PMP EoS  (\ref{EqIX})
in such a way that it does not contradict to QCD phenomenology. 
 
\label{sec:conc}
\section*{Conclusions}

In this work we perform  a thorough  inspection of the irregularities of thermodynamic 
quantities  observed at CFO and discuss the possible signals of  two QCD phase transitions.  
The most remarkable irregularities include 
 two sets of  correlated  quasi-plateaus found in  \cite{Bugaev:2014,Bugaev:2015}  which are  located at the  collision energy ranges
$\sqrt{s_{NN}} \simeq 3.8-4.9$ GeV  and  $\sqrt{s_{NN}} \simeq  7.6-9.2$ GeV, and two  peaks 
of trace anomaly $\delta$ observed at the maximal energy  of  each set of  quasi-plateaus. 
Using the most advanced version of the hadron resonance gas model which allows one to safely go beyond the  standard  Van der Waals 
approximation, here we found two strong  peaks of the baryonic charge  density located exactly at the collision energies of the trace anomaly peaks, i.e.  at  $\sqrt{s_{NN}} = 4.9$ GeV and $\sqrt{s_{NN}} = 9.2$ GeV.  The usage of the NHRGM  which  is not restricted by the Van der Waals approximation was necessary, since at the mentioned peaks  the baryonic charge density   is  high  and  corresponds to   $1.25-1.4$ values of the normal nuclear density $\rho_0 = 0.16$~fm$^{-3}$.  Moreover, during the fitting even higher values of baryonic charge density  were analyzed.  

In addition in this work we closely  studied the 
collision energy  dependence of the ratio of  total number of strange quarks and antiquarks to the  number of non-strange quarks and antiquarks  (a modified Wroblewski factor $\lambda_s$)  and the strangeness  suppression factor $\gamma_s$. 
With the help of NHRGM we find   a sizable  jump of the $\lambda_s$ factor, when the collision energy increases from   $\sqrt{s_{NN}} = 4.3$ GeV to $\sqrt{s_{NN}} = 4.9$ GeV, while at  the collision energies above  $\sqrt{s_{NN}} = 7.6$ GeV we observe    a  saturation  of  the $\lambda_s$ factor.  Similarly  to previous findings  \cite{SFO,Veta14,Sagun17}  we  observe rather complicated  energy dependence of    $\gamma_s$ factor: 
at  $\sqrt{s_{NN}} = 3.8$ GeV and $\sqrt{s_{NN}} = 7.6$ GeV 
there exist two maxima, while at   $\sqrt{s_{NN}} = 4.9$ GeV and at  $\sqrt{s_{NN}} >  8.7$ GeV the  strangeness 
seems to reach  a chemical equilibrium. 

Since  the low energy set of quasi-plateaus along with  the trace anomaly  peak  at   $\sqrt{s_{NN}} = 4.9$ GeV are unambiguously  explained  by   the anomalous thermodynamic properties of  mixed phase  of  the 1-st order phase transition formed at    
$\sqrt{s_{NN}} = 4.3-4.9$ GeV \cite{Bugaev:2014,Bugaev:2015}, we conclude  that other irregularities observed at   $\sqrt{s_{NN}} = 4.9$ GeV, i.e. the peak of baryonic charge density, the $\lambda_s$ factor  jump and the strangeness equilibration  can be the other signals of this  phase transition. 
In this work we give a straightforward  explanation to the  strangeness equilibration  at   $\sqrt{s_{NN}} = 4.9$ GeV  and at 
$\sqrt{s_{NN}} >  8.7$ GeV.  Based on the unique thermostatic properties of  Hagedorn mass spectrum we  follow Ref. \cite{Sagun17} and   suggest that  at $\sqrt{s_{NN}} >  8.7$ GeV  the formation of  quark-gluon bags with the exponential mass spectrum  forces the hadrons born at  the moment of hadronization to be in a full thermal and chemical equilibrium  with the bags.  
Similarly, in this work we notice that  under the condition of the constant external pressure any mixed phase of the 1-st order 
phase transition is an explicit thermostat and explicit particle reservoir and, hence, anything born out of it should be 
in a full thermal and  chemical equilibrium with the mixed phase.  
Since the mixed phase   EoS  employed in 
Refs.   \cite{Bugaev:2014,Bugaev:2015}, indeed, leads to  an approximate collision  energy independence of  initial pressure of a  system formed in the collision process, we came to a natural conclusion that  the formation of mixed phase  at   $\sqrt{s_{NN}} = 4.9$ GeV 
is responsible for the chemical equilibration of  strangeness  at this  energy.  Thus, the  explicit thermostatic properties of the mixed phase cause  an appearance of  the  strangeness equilibrium dale at  $\sqrt{s_{NN}} = 4.9$ GeV. On the other hand, due to an absence of an alternative explanation of such a  dale, its appearance, in turn, is an independent and strong argument in favor of the mixed phase formation at this collision energy. 

As it is argued here and in Refs.  \cite{Bugaev:2014,Bugaev:2015} the above mentioned irregularities observed at the collision 
energy range  $\sqrt{s_{NN}} \simeq 3.8-4.9$ GeV are the signals of the 1-st order phase transition, then, according to our hypothesis,  a set of similar irregularities found at the collision energies  $\sqrt{s_{NN}} \simeq  7.6-9.2$ GeV should be associated 
with another phase transition. Such a hypothesis is supported by the detailed analysis of  $K^+/\pi^+$ ratio of  Ref.   \cite{Nayak2010} and by the one of light nuclei fluctuations of Ref.  \cite{Ko2017}.  In addition to  thermodynamic and hydrodynamic signals of phase transition  discussed here the work  \cite{Ko2017}  provides us with the fluctuation signal
of the 2-nd order phase transition at the vicinity of   $\sqrt{s_{NN}} \simeq 8.8-9.2$ GeV.  Therefore, we argue that 
the collision energy range  $\sqrt{s_{NN}} \simeq 8.8-9.2$ GeV may be the nearest vicinity of  3CEP.

The EoS (\ref{EqIX}) used  in  \cite{Bugaev:2014,Bugaev:2015}  to model  the properties of high density matter  formed 
at the initial stage of collision for the energies  $\sqrt{s_{NN}} \simeq 6.3-9.2$ GeV  provides  us with  a unique opportunity to
reveal the total number of  massless bosonic and fermionic degrees of freedom.  The found  number is about 1770, which is  too large for elementary degrees of freedom of  QCD, but it  has the same order of magnitude  as the total number of experimentally known hadronic spin-isospin states.    {Although, at the moment there is insufficient data to  discuss a possible 
origin of  this phase,  our hypothesis that  such a phase corresponds to a gas  of  interacting,  but   almost massless hadrons,  is strongly supported by the results of  PHSD microscopic model  which accounts for the effects of 
hadronic and partonic mass reduction in a very similar energy range \cite {Cassing16a,Cassing16b}}. 
Therefore,   we explicitly show that, if one accounts for the  excluded volume of  these hadronic degrees of freedom in a relativistic fashion,  then at high energy densities the EoS (\ref{EqIX}) can be   modified in such a way that it does not contradict  to 
the QCD thermodynamics. 

The above results and  hypotheses are rather speculative, but they reflect  the present state of art of heavy ion collision phenomenology, where at the moment  there is insufficient experimental and theoretical  information right  at the most promising regions of  the QCD phase diagram. Therefore, we hope that  this work will stimulate both theoreticians and experimentalists 
to study the discussed  problems in more details  in order to  prove or disprove our results. 

\vspace*{2.8mm}
\noindent 
{\bf Acknowledgments.}
The authors appreciate the valuable comments of  E. E. Kolomeitsev, S. N. Nedelko and A. I. Titov.
K.A.B., A.I.I., V.V.S. and G.M.Z.  acknowledge  a partial  support from 
the program ``Nuclear matter under extreme conditions'' launched 
by the Section of Nuclear Physics of  the National Academy of Sciences  of Ukraine. 
 The work of K.A.B. and L.V.B. was performed in the framework of COST Action CA15213 ``Theory of hot matter and relativistic heavy-ion collisions" (THOR). K.A.B. is thankful to the  COST Action CA15213  for a partial support. 
 V.V.S. acknowledges a partial support by grants from ``Funda\c c\~ao para a Ci\^encia e Tecnologia".
The work of D.B.  was supported in part by the Polish National Centre (NCN) under contract number 
UMO-2011/02/A/ST2/00306.  The work  of A.V.T. was partially supported by the Ministry of Science and Education
of the Russian Federation, grant No 3.3380.2017/4.6, and by  National
Research Nuclear University ``MEPhI'' in the framework of the Russian
Academic Excellence Project (contract No 02.a03.21.0005, 27.08.2013).



\end{document}